\documentstyle[12pt,epsfig]{article}
\def\be{\begin{equation}} \def\ee{\end{equation}} \def\bea{\begin{eqnarray}}
\def\eea{\end{eqnarray}} \def\nnb{\nonumber}

\begin{document}

\hfill{March 21, 2016}

\begin{center}
\vskip 6mm 
\noindent
{\Large\bf  
Elastic $\alpha$-$^{12}$C scattering at low energies 
in cluster effective field theory
}
\vskip 5mm 

\noindent
{\large 
Shung-Ichi Ando\footnote{mailto:sando@sunmoon.ac.kr}, 
\vskip 5mm
\noindent
{\it
School of Mechanical and ICT Convergence Engineering,
\\
Sunmoon University,
Asan, Chungnam 31460,
Republic of Korea
}
}
\end{center}

\vskip 5mm

The elastic $\alpha$-$^{12}$C scattering 
at low energies is studied employing an effective field theory
in which the $\alpha$ and $^{12}$C states are treated 
as elementary-like fields.
We discuss scales of the theory at stellar energy region 
that the ${}^{12}$C($\alpha$, $\gamma$)$^{16}$O process occurs, 
and then obtain an expression of the elastic scattering amplitudes 
in terms of effective range parameters.
Using experimental data of the phase shifts for $l=0,1,2$ channels
at low energies, 
for which the resonance regions are avoided,
we fix values of the parameters
and find that the phase shifts at the low energies are well reproduced
by using three effective range parameters for each channel. 
Furthermore, we discuss problems and uncertainties of the present approach
when the amplitudes are extrapolated to the stellar energy region.

\vskip 5mm 
\noindent
PACS(s): 
11.10.EF, 
24.10.-i, 
25.55.Ci, 
26.20.Fj. 

\newpage
\vskip 2mm \noindent
{\bf 1. Introduction}

The radiative alpha capture on carbon-12, 
${}^{12}$C($\alpha$, $\gamma$)${}^{16}$O, 
is one of the fundamental reactions in 
nuclear astrophysics, which determines
the ratio ${}^{12}$C/${}^{16}$O produced in helium burning~\cite{f-rmp84}.
The reaction rate, equivalently the astrophysical $S$-factor, of
the process at the Gamow peak energy, $T_G=0.3$~MeV, however, cannot
experimentally be determined due to the Coulomb barrier. 
It is necessary to employ a theoretical model and extrapolate 
the cross section down to $T_G$ 
by fitting the model parameters to available experimental data 
measured at a few MeV or larger.
During a last half century, a lot of experimental and theoretical 
studies for the process have been carried out.
For reviews, see, e.g., Refs.~\cite{bb-npa06,chk-epja15} 
and references therein. 

In constructing a model for the process, one needs to take account of 
excited states of ${}^{16}$O~\cite{bb-npa06}, 
particularly, two excited bound states for 
$l^\pi_{n\mbox{-}th}=1_1^-$ and $2_1^+$ 
just below the $\alpha$-${}^{12}$C breakup threshold at
$T=-0.045$ and $-0.24$~MeV~\footnote{
The energy $T$ denotes that of the $\alpha$-${}^{12}$C system 
in center of mass frame. 
}, respectively, 
as well as $1_2^-$ and $2_2^+$ resonant (second excited) states at 
$T=2.42$ and $2.68$~MeV, respectively.
Thus the capture reaction to the ground state of ${}^{16}$O 
at $T_G$ is expected to be E1 and E2 transition dominant 
due to the subthreshold $1_1^-$ and $2_1^+$ states. 
While the resonant $1_2^-$ and $2_2^+$ states play 
a dominant role in the available
experimental data at low energies, typically $1\le T \le 3$~MeV. 
Experimental data pertaining to processes for nuclear astrophysics 
are compiled, 
known as NACRE-II compilation~\cite{NACRE-II},
in which the $S$-factor of 
the ${}^{12}$C($\alpha$,$\gamma$)${}^{16}$O reaction
is estimated employing a potential model 
and reported uncertainty of the process is less than 20~\%. 
While conflicting sets of experimental data for the process
at very low energies still persist~\cite{g-jpg98,g-prc13}, 
and thus one may need to wait for new measurements at very low energies, 
$T\le 1.5$~MeV~\cite{chk-epja15}.  

In the present study, 
we would like to discuss an alternative theoretical approach
constructing an effective field theory (EFT) for the process, 
and apply the theory to the study of elastic $\alpha$-${}^{12}$C
scattering at low energies. 
EFTs provide us a model independent and systematic method
for theoretical calculations.
An EFT for a system in question can be built by introducing 
a scale which separates relevant degrees of freedom at low energies
from irrelevant degrees of freedom at high energies. 
An effective Lagrangian is written down in terms of the relevant degrees 
of freedom,
and is perturbatively expanded 
order by order, 
by counting the number of derivatives. 
The irrelevant degrees of freedom are integrated out
and their effect is embedded in coefficients appearing in the effective
Lagrangian. 
Thus a transition amplitude is systematically
calculated by writing down Feynman diagrams, 
whereas the coefficients appearing in the effective Lagrangian 
should be determined by experiments. 
For reviews, one may refer to, e.g., 
Refs.~\cite{bv-arnps02,bh-pr06,m-15}.
Various processes being essential in nuclear astrophysics 
have been investigated
by constructing EFTs, for example, 
$p(n,\gamma)d$ at BBN energies~\cite{r-npa00,achh-prc06}
and $pp$ fusion~\cite{kr-npa99,bc-plb01,aetal-plb08,cly-plb13} 
and ${}^7$Be($p$,$\gamma$)${}^7$B~\cite{znp-prc14,rfhp-epja14}
in the Sun. 

An unique feature of those studies in EFTs is that the theories
allow us to estimate {\it theoretical} uncertainties, based on the 
model-independent and perturbative expansion scheme of the theories,
in the extrapolated reaction rates. For example, less than 1~\% 
accuracy in the estimate of the reaction rates of $p(n,\gamma)d$
at BBN energies and the $pp$ fusion in the Sun were obtained 
in the previous studies~\cite{achh-prc06,aetal-plb08}.
Thus our main aim in future studies for 
the $^{12}$C($\alpha$,$\gamma$)$^{16}$O
reaction is to estimate the $S$ factors with about 5~\%
uncertainty in theory.

We treat the $\alpha$ and ${}^{12}$C states as elementary-like 
fields, and the scales involving in the theory are to be discussed in 
the next section.
Then an effective Lagrangian is written down, 
an expression of the scattering amplitudes is obtained, and 
phase shifts for $l=0,1,2$ channels 
of the elastic $\alpha$-${}^{12}$C scattering
at low energies are studied.
The main assumption of the present study, 
suggested by Teichmann~\cite{t-pr51}, 
is that we may choose the energies of the resonant states 
the large energy scale of the theory 
so that, in the limited low energy regions, the Breit-Wigner type
pole structure for the resonant states in the scattering amplitudes can be 
expanded in terms of the energy and the expression of the energy 
dependence of the amplitudes can be reduced to 
that of the effective range expansion. 
Thus our large energy scales of the theory for the elastic scattering
for $l=0,1,$ and 2 channels
are the resonance energies, $T=4.89, 2.42,$ and 2.68~MeV 
for the $0_2^+$, $1^-_2$, $2_2^+$ states, respectively.
In addition, we do not introduce explicate degrees of freedom
for the $1_1^-$ and $2^+_1$ states.
Because, as to be discussed in detail later, 
the expression of the scattering amplitudes 
in terms of the effective range parameters
have a restrictive condition in zero momentum limit,
we find that it is not easy to incorporate 
the subthreshold states in the present study.

This article is organized in the following.
In section 2, 
we discuss the scales of the theory and write
down an effective Lagrangian for the elementary-like
$\alpha$ and ${}^{12}$C fields.
In section 3, the expression of the amplitudes
for the elastic $\alpha$-${}^{12}$C scattering for $l=0,1,2$
channels in terms of the effective range parameters
is obtained. 
In section 4, the parameters are fitted by using the experimental
phase shifts, and for a qualitative study of the extrapolation
the real part of the denominator of the scattering 
amplitudes is extrapolated to $T_G$. 
Finally, conclusions and discussion of the present
work are presented in section 6.

\vskip 2mm \noindent
{\bf 2. Scales and effective Lagrangian for the system}

As mentioned above, we treat the $\alpha$ and ${}^{12}$C states
as elementary-like cluster fields. 
This treatment would be reasonable
when a typical momentum scale 
is smaller than a scale at which a mechanism at high energy 
becomes relevant.
For the $\alpha$ particle, excited states of the $\alpha$ particle 
should be treated as irrelevant degrees 
of freedom~\cite{hhvk-npa08,ao-prc14}.
First excited energy of the $\alpha$ particle is
$E_{(4)}\simeq 20$~MeV, 
and thus a corresponding large momentum scale is 
$\Lambda_H \sim \sqrt{2\mu_4 E_{(4)}} \simeq 170$~MeV  
where $\mu_4$ is the reduced
mass 
for one and three nucleon systems,
$\mu_4\simeq \frac34m_N$. $m_N$ is the nucleon mass.
For the $^{12}$C state, 
on the other hand,
first excited energy of ${}^{12}$C is $E_{(12)}\simeq 4.439$~MeV,
and thus the large momentum scale due to 
$E_{(12)}$
is $\Lambda_H \sim \sqrt{2\mu_{12}E_{(12)}}\simeq 150$~MeV 
where $\mu_{12}$ is 
the reduced mass for four and eight nucleon systems,
$\mu_{12}\simeq \frac83m_N$.
In addition, we need to introduce another large scale due to 
the Coulomb interaction.
The inverse of the Bohr radius is 
$\kappa=\alpha_E Z_\alpha Z_C \mu\simeq 247$~MeV 
where $\alpha_E$ is the fine structure constant, 
$Z_\alpha$ and $Z_C$ are the number of protons in 
$\alpha$ and ${}^{12}$C, respectively, 
and $\mu$ is the reduced mass for
$\alpha$ and ${}^{12}$C, $\mu\simeq 3m_N$. 
Thus we may choose the large momentum scale of the theory
$\Lambda_H\sim 150$~MeV. 

A typical momentum scale $Q$ for 
the ${}^{12}$C($\alpha$,$\gamma$)${}^{16}$O process
in the starts is estimated from the Gamow peak energy, 
$T_G\simeq 0.3$~MeV, and thus the typical momentum scale is 
$Q \sim k = \sqrt{2\mu\, T_G}\simeq 41$~MeV.
Thus we shall have the expansion parameter for the process at
$T_G$ as $Q/\Lambda_H\sim 1/3$,
and the about 5~\% theoretical uncertainty mentioned above 
can be achieved by considering perturbative corrections 
up to next-to-next-to leading order. 

A typical momentum scale, $Q\sim k$, 
for the elastic $\alpha$-${}^{12}$C scattering 
differs from that at $T_G$. 
We employ the phase shift
data from Plaga {\it et al.}~\cite{petal-npa87} and 
Tischhauser {\it et al.}~\cite{tetal-prc09} to fix the effective
range parameters. 
The reported energies of the $\alpha$
particle for the phase shift data 
in lab frame are $T_\alpha \simeq$ 1.5-6.6~MeV and 
2.6-6.6~MeV\footnote{
The energies $T_\alpha$ and $T$ for the $\alpha$-${}^{12}$C system 
in lab and center of mass frames are related 
by $T_\alpha\simeq \frac43 T$.
}, respectively, whereas, as mentioned above,
we introduced the resonance energies as the large scales 
of the process. 
Thus we have the lowest momenta in the center of mass frame,
$k_{low}\simeq 80$ and 105~MeV for Ref.~\cite{petal-npa87}
and \cite{tetal-prc09}, respectively, whereas the highest momenta,
$k_{high} \simeq 166, 117,$ and 123~MeV for $l=0,1,$ and 2 channels,
respectively. 
Because the large momentum scale of the theory is $\Lambda_H\sim 150$~MeV, 
though in the higher momentum region the series expansion would not
converge, it may do in the relatively low momentum region. 
The convergence of the effective range expansion for 
each channel is to be studied below.

An effective Lagrangian for the present study 
may be written as~\cite{hhvk-npa08,bhvk-plb03,ah-prc05}
\bea
{\cal L} &=& \phi_\alpha^\dagger \left(
iD_0 
+\frac{\vec{D}^2}{2m_\alpha}
+ \cdots
\right) \phi_\alpha
+ \phi_C^\dagger\left(
iD_0
+ \frac{\vec{D}^2}{2m_C}
+\cdots
\right)\phi_C
\nnb \\ && +
\sum_{l,n}C_{n}^{(l)}d_{(l)}^\dagger \left[
iD_0 
+ \frac{\vec{D}^2}{2(m_\alpha+m_C)}
\right]^n d_{(l)}
\nnb \\ && 
- \sum_{l}y_{(l)}\left[
(\phi_\alpha O_l \phi_C)^\dagger d_{(l)}
+ d_{(l)}^\dagger(\phi_\alpha O_l \phi_C)
\right]
+ \cdots\,,
\eea
where $\phi_\alpha$ ($m_\alpha$) and 
$\phi_C$ ($m_C$) are point-like fields (masses) of $\alpha$ and $^{12}$C, 
respectively. 
$D_\mu$ is a covariant derivative, and
the dots denote higher order terms.
$d_{(l)}$ represent
$\alpha$ and $^{12}$C composite fields of angular momentum $l$.
Thus $d_{(0)}$ for $l=0$, $d_{(1)i}$ for $l=1$ 
where the subscript $i$ represents a state in $l=1$, and $d_{(2)ij}$ 
for $l=2$ where $d_{(2)ij}=d_{(2)ji}$ and the subscripts $ij$ represent
a state in $l=2$.
$C_{n}^{(l)}$ are coupling constants for 
the propagation of the $\alpha$-$^{12}$C composite fields 
for the $l$ channels,
and can be related to effective range parameters 
along with common multiplicative factors $1/y_{(l)}^2$. 
For the present exploratory study, three effective range parameters, 
the terms of $n=0,1,2$, are retained 
for each partial wave.
For the $l=0$ state, for example, $C_{0}^{(0)}$ is related to 
the scattering length, $C_{1}^{(0)}$ the effective range, and $C_{2}^{(0)}$
the shape parameter. 
In addition, 
$y_{(l)}$ are coupling constants of the $\alpha$-$^{12}$C-$d_{(l)}$ vertices, 
and $O_l$ are projection operators
by which the $\alpha$-$^{12}$C system is projected to the $l$-th 
partial wave states. Thus one
has
\bea
O_0 = 1\,,
\ \ \ 
O_{1,i} = 
\frac{i\stackrel{\leftrightarrow}{D}_i}{M} 
\equiv 
i \left(
\frac{\stackrel{\rightarrow}{D}_C}{m_C} -
\frac{\stackrel{\leftarrow}{D}_\alpha}{m_\alpha} 
\right)_i
\,, 
\ \ \ 
O_{2,ij} = 
\frac{1}{M^2}\left(
- \stackrel{\leftrightarrow}{D}_i
\stackrel{\leftrightarrow}{D}_j
+ \frac13 \delta_{ij}
\stackrel{\leftrightarrow}{D}^2
\right) \,.
\eea

\vskip 2mm \noindent
{\bf 3. Scattering amplitudes and phase shifts}

The differential cross section of the elastic $\alpha$-$^{12}$C scattering 
(for two spin-0 charged particles)
in terms of the phase shifts are given by
(see, e.g., Ref.~\cite{l-rmp57})
\bea
\sigma(\theta) &=&  
\frac{d\sigma}{d\Omega} = |f(\theta)|^2
\nnb \\ &=&
\frac{1}{k^2} \left|
-\frac{\eta}{2\sin^2\frac12\theta}
\exp\left(-2i\eta \ln\sin\frac12\theta\right)
\right.  \nnb \\ && \left.
+ \frac12 i \sum_{l=0}^\infty (2l+1) \left[
\exp\left(2i\omega_l\right)
-U_l\right] P_l(\cos\theta)
\right|^2\,,
\label{eq;sigma-theta}
\eea
where $f(\theta)$ is the scattering amplitude
including both pure Coulomb part 
and Coulomb modified strong interaction part, 
$\theta$ is a scattering angle, 
$k$ is the relative absolute momentum,
and
$\eta = \kappa/k$. 
In addition,
$\omega_l$ is 
the Coulomb scattering phase, 
$\omega_l (= \sigma_l-\sigma_0) 
= \sum \arctan(\eta/s)$ for $s=1$ to $l$,\footnote{
$\sigma_0$ is the Coulomb phase shift, 
$\sigma_0=\arg\Gamma(1+i\eta)$.
}
and 
\bea
U_l = \exp\left[
2i(\delta_l + \omega_l)\right]\,.
\eea
$\delta_l$ are real scattering phase shifts.

\begin{figure}
\begin{center}
\includegraphics[width=12cm]{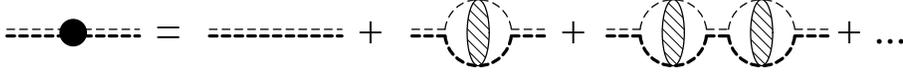}
\caption{
Diagrams for propagator of dressed composite fields.
Double dashed line denotes bare composite (${}^{16}$O) 
field consisting of $\alpha$ and ${}^{12}$C fields,
thin (thick) dashed line denotes point-like $\alpha$ (${}^{12}$C)
field, and shaded blob denotes off-shell Coulomb T-matrix. 
}
\label{fig;propagator}
\end{center}
\end{figure}

\begin{figure}
\begin{center}
\includegraphics[width=3.5cm]{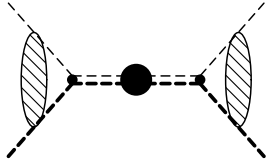}
\caption{
Diagram for scattering amplitudes. 
See the caption in Fig.~\ref{fig;propagator} as well.
}
\label{fig;amplitude}
\end{center}
\end{figure}

The elastic scattering amplitudes for the Coulomb modified strong 
interaction part for $l=0,1,2$ channels are calculated 
from the effective Lagrangian presented above.
In Fig.~\ref{fig;propagator} Feynman diagrams for dressed 
composite ${}^{16}$O propagators consisting of the $\alpha$ and ${}^{12}$C
elementary-like fields including the Coulomb interaction between
the two charged fields 
are depicted. 
In Fig.~\ref{fig;amplitude}, a Feynman diagram for a scattering amplitude
for elastic $\alpha$-${}^{12}$C scattering for each partial wave state 
including the initial and final state Coulomb interactions is depicted.
For derivation of the amplitudes from the diagrams in detail
the reader may referee to, e.g., Refs.~\cite{ashh-prc07,a-epja07} 
and we do not repeat the detailed calculation.

Thus we have the scattering amplitudes, $A_l$,  
for $l=0,1,2$ states in terms of the effective range parameters 
as~\cite{hhvk-npa08,h-epjwc10}
\bea
A_0 &=& \frac{2\pi}{\mu}
\frac{C_\eta^2 e^{2i\sigma_0}}{
-\gamma_0
+ \frac12 r_0 k^2 - \frac14P_0 k^4 
-2\kappa H(\eta)}\,,
\label{eq;A0}
\\
A_1 &=& \frac{6\pi}{\mu}
\frac{e^{2i\sigma_1}k^2(1+\eta^2)C_\eta^2 \cos\theta}{
-\gamma_1
+ \frac12 r_1 k^2
-\frac14 P_1 k^4
-2\kappa k^2(1+\eta^2) H(\eta)}\,,
\label{eq;A1}
\\ 
A_2 &=& \frac{10\pi}{\mu}
\frac{e^{2i\sigma_2}k^4(4+\eta^2)(1+\eta^2)C_\eta^2
\frac12(3\cos\theta -1)}{
-\gamma_2
+ \frac12 r_2 k^2
- \frac14 P_2 k^4
-2\kappa k^4(4+\eta^2)(1+\eta^2)H(\eta)}\,,
\label{eq;A2}
\eea
with
\bea
C_\eta^2 = \frac{2\pi\eta}{e^{2\pi\eta} -1}\,,
\ \ \ \
H(\eta) = \psi(i\eta) + \frac{1}{2i\eta} - \ln(i\eta)\,,
\eea
where $\psi(x)$ is the digamma function.
$\gamma_l$, $r_l$, $P_l$ are the three effective range parameters
for $l=0,1,2$.
While the amplitudes, $A_l$, can be represented 
in terms of the phase shifts $\delta_l$ as
\bea 
A_0 =  \frac{2\pi}{\mu}\frac{e^{2i\sigma_0}}{
k\cot\delta_0 -ik}\,,
\ \ \
A_1 
= \frac{6\pi}{\mu}
\frac{e^{2i\sigma_1}\sin\theta}{
k\cot\delta_1
-ik}\,,
\ \ \
A_2 = \frac{10\pi}{\mu}
\frac{e^{2i\sigma_2}\frac12(3\cos\theta-1)}{
k\cot\delta_2 -ik}\,.
\eea
Thus one has the relations between the phase shifts and 
the effective range parameters as 
\bea
C_\eta^2 k \cot\delta_0
+ 2\kappa h(\eta) &=&
-\gamma_0
+ \frac12 r_0 k^2
- \frac14 P_0 k^4
+ \cdots\,,
\label{eq;l=0}
\\ 
k^2(1+\eta^2)\left[
C_\eta^2 k \cot\delta_1
+ 2\kappa h(\eta)
\right] &=&
-\gamma_1
+ \frac12 r_1 k^2
- \frac14 P_1 k^4
+ \cdots\,,
\label{eq;l=1}
\\ 
k^4(4+\eta^2)(1+\eta^2)\left[
C_\eta^2 k \cot\delta_2
+ 2\kappa h(\eta)
\right] &=&
-\gamma_2
+ \frac12 r_2 k^2
- \frac14 P_2 k^4
+ \cdots\,,
\label{eq;l=2}
\eea
where $h(\eta) = ReH(\eta)$.

\vskip 2mm \noindent
{\bf 4. Fixing the parameters}

Before fixing values of the effective range parameters,
we discuss some features of the equations obtained 
in Eqs.~(\ref{eq;l=0},\ref{eq;l=1},\ref{eq;l=2}).
At low energies 
the function $h(\eta)$ 
appearing in the equations
can be expanded in terms of
1/$\eta (=k/\kappa)$ as 
\bea
h(\eta) = \frac{1}{12\eta^2} 
+ \frac{1}{120\eta^4} 
+ \frac{1}{252\eta^6} 
+ \cdots\,.
\eea
One may see that 
the series expansion converges in the energy region which 
we consider below and  
there is no constant term appearing from the $h(\eta)$ 
function. 
In addition, the factor $C_\eta^2$ 
being multiplied to the cotangent terms 
in Eqs.~(\ref{eq;l=0},\ref{eq;l=1},\ref{eq;l=2})
becomes vanishingly small at the very low energies. 
Thus the left hand side of the equations vanishes in 
zero momentum limit, $k\to 0$, and
the parameters $\gamma_0$, $\gamma_1$, and $\gamma_2$ 
in the right hand side of
 Eqs.~(\ref{eq;l=0}), (\ref{eq;l=1}), and (\ref{eq;l=2}) are required 
to vanish as well in the limit. 
On the other hand, 
experimental data do not exist at such very low energies,
and values of the parameters are fixed by
using existing experimental data
at some higher energies. 
As mentioned above, the experimental data 
from Plaga {\it et al.}~\cite{petal-npa87} and 
Tischhauser {\it et al.}\cite{tetal-prc09} where the lowest energies 
the data are $T_\alpha \simeq 1.5$ and 2.6~MeV, respectively 
are employed up to the energies of the resonant states,
$T_\alpha \simeq 6.5$, 3.2, and 3.6~MeV for 
$l=0$, 1, and 2, respectively.\footnote{
Thus the momentum of the $\alpha$-${}^{12}$C system in the center of 
mass frame for the data becomes $k=80(105)$-166~MeV for $l=0$,
80(105)-117~MeV for $l=1$, and 80(105)-123~MeV for $l=2$.
}

We note that we have to choose $\gamma_2=0$ in fitting the parameters
because the phase shift for the $l=2$ channel is very small, 
less than two degrees, 
in the fitting energy range and it is not easy to 
have a non-vanishing contribution to $\gamma_2$. 
In addition, due to the feature in the zero momentum limit
mentioned above, it is not easy to incorporate the pole structure
of the subthreshold states in the amplitudes either
because it makes the $\gamma_l$ terms significantly large.
Therefore, we fix the parameters,
without including the pole structure of the subthreshold states,
from data sets which we arbitrarily choose 
for each of the partial wave states, $l=0,1,2$, below.

\vskip 2mm \noindent
{\bf 4.1. $l=0$ channel}

\begin{table}[t]
\begin{center}
\begin{tabular}{c|c c c c}
               &
$S0$ & $S1$ & $S2$ & $S3$ \\ \hline 
$\gamma_0$ (MeV) & 
 $0.058 \pm 0.058$ & $0.034\pm 0.003$ & $0.015\pm0.001$ & $-0.008\pm0.001$ \\
$r_0$ (fm) &
 $0.270\pm 0.002$ & $0.2693\pm 0.0001$ & $0.2685\pm 0.0001$ & 
 $0.2674\pm 0.0000$ \\
$P_0$ (fm$^3$) &
 $-0.037\pm0.005$ & $-0.0372\pm0.0002$ & $-0.0390\pm0.0001$ & $-0.0416\pm0.0000$
\\ \hline 
\end{tabular}
\caption{
Fitted values of $s$-wave effective range parameters using four 
sets of the experimental data labeled by $S0$, $S1$, $S2$, and $S3$.
See the text for details.
}
\label{table;s-wave_effective_ranges}
\end{center}
\end{table}

Four sets of the experimental data
for the $s$-wave phase shift are chosen
in order to qualitatively study the dependence 
from the choice of the data sets
for the extrapolation to $T_G$.
The four sets of the data are labeled as 
$S0$, $S1$, $S2$, and $S3$. 
$S0$ denotes a data set of the $s$-wave phase shift at
energies $T_\alpha = 1.5$-6.5~MeV from Table 2 
in the Plaga {\it et al.}'s paper~\cite{petal-npa87}, and 
$S1$, $S2$, and $S3$ do those 
at energies $T_\alpha = 2.6$-6.5, 2.6-6.0, and 2.6-5.0~MeV, respectively, 
from the Tischhauser {\it et al.}'s paper~\cite{tetal-prc09}. 

In Table \ref{table;s-wave_effective_ranges} fitted values 
of the $s$-wave effective range parameters by using the four
data sets, $S0$, $S1$, $S2$, and $S3$ are 
displayed.\footnote{
We employ a scipy module, {\tt curve\_fit},
in optimization package to fit the effective range parameters 
to the phase shift data.
} 
One can see that the fitted values of $\gamma_0$ are sensitive 
to the choice of the data sets, those of $r_0$ are not,
and those of $P_0$ are in between the two cases. 
We find that almost 
exact cancellations between the $r_0$ term and the coefficient of
the term proportional to $k^2$, $1/(3\kappa)\simeq 0.2687$~fm, 
from $2\kappa h(\eta)$ term in Eq.~(\ref{eq;l=0}) and 
significant cancellations between the $P_0$ term and that of the term 
proportional to $k^4$ term, $-1/(15\kappa^3)\simeq$ $-0.0210$~fm$^3$, 
from the $2\kappa h(\eta)$ term.  
As discussed in the introduction, we find that 
the expansion series in terms of the effective range parameters 
well converges in the energy regions for the fitting.
Those coefficients 
of the $k^{2n}$ power series after including the corrections 
from the $2\kappa h(\eta)$ term
become significantly small, 
e.g., 
the $\gamma_0$ values being a few hundredth MeV,
compared to the scale of the system.
This may be due to the suppression factor from the $C_\eta^2$ term
in Eq.~(\ref{eq;l=0}),
which becomes $C_\eta^2\sim 10^{-6}$-$10^{-4}$ in the range of 
the energy, $T_\alpha\simeq 2.0$-$6.0$~MeV.

\begin{figure}[th]
\begin{center}
\includegraphics[width=12cm]{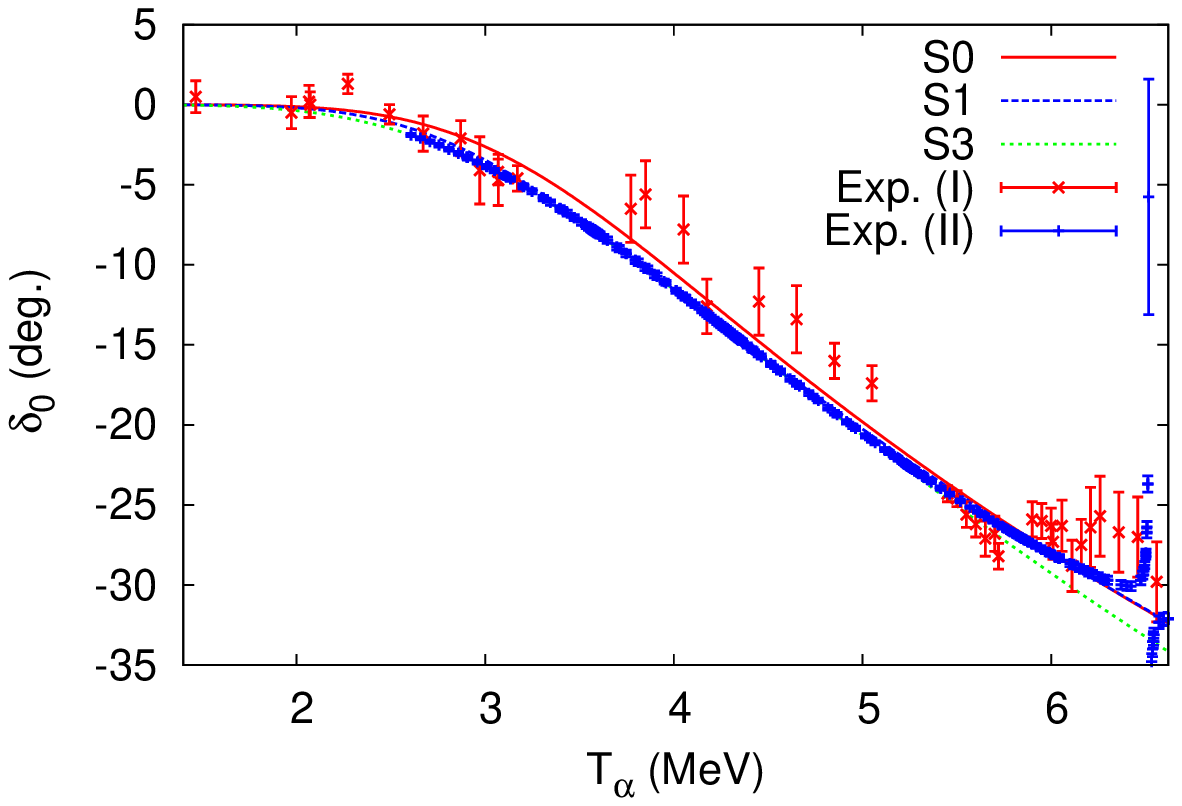}
\caption{
Phase shift of elastic $\alpha$-$^{12}$C scattering for $l=0$, 
$\delta_0$ (deg.), as functions of $T_\alpha$ (MeV).
Three curves are plotted by using three sets of fitted effective range 
parameters (labeled by $S0$, $S1$, $S3$) obtained 
in Table~\ref{table;s-wave_effective_ranges}. 
Experimental data labeled by
Exp. (I) from Plaga {\it et al.}~\cite{petal-npa87} 
and Exp. (II) from Tischhauser {\it et al.}~\cite{tetal-prc09}
are also displayed.
}
\label{fig;phaseshift-sw}
\end{center}
\end{figure}
\begin{figure}[th]
\begin{center}
\includegraphics[width=12cm]{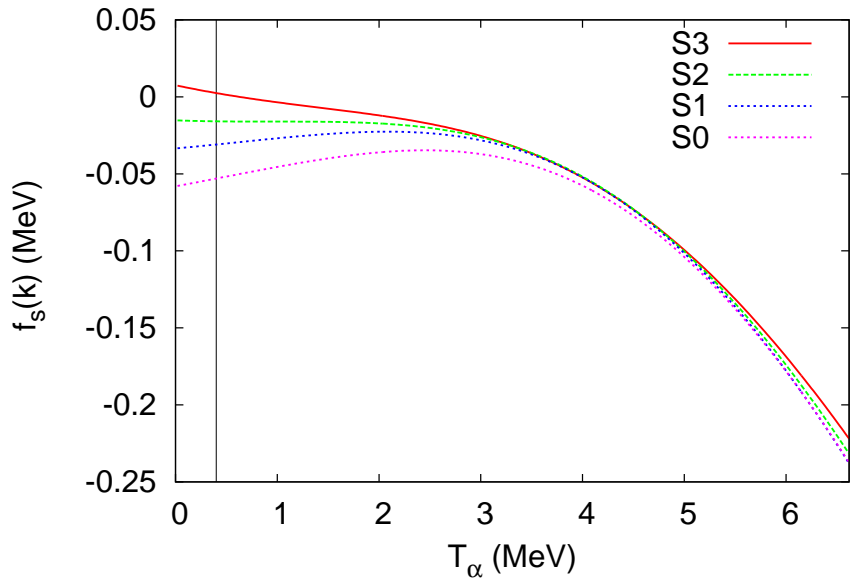}
\caption{
Function $f_s(k)$ defined in Eq.~(\ref{eq;fs}) as functions of $T_\alpha$ (MeV).
Curves are plotted 
by using four sets of values of effective range parameters (labeled 
by $S0$, $S1$, $S2$, $S3$) obtained 
in Table \ref{table;s-wave_effective_ranges}. 
A vertical line at $T_\alpha=0.4$~MeV is also included.
}
\label{fig;fx-sw}
\end{center}
\end{figure}

In Fig.~\ref{fig;phaseshift-sw}, curves of the $s$-wave phase shift
are plotted
by using the effective range parameters obtained
in Table~\ref{table;s-wave_effective_ranges}.
The experimental data are also included in the figure.
We find that the curves are well reproduced the data in the 
energy ranges 
where the effective range parameters have been fitted.

In Fig.~\ref{fig;fx-sw}, in order
to qualitatively study the extrapolation to the Gamow energy, $T_G=0.3$~MeV
in the center of mass frame,
which corresponds to $T_\alpha\simeq 0.4$~MeV in the lab frame,
we plot curves of the real part of the denominator of 
the $s$-wave scattering amplitude in Eq.~(\ref{eq;A0}),
\bea
f_s(k) = - 2\kappa h(\eta) - \gamma_0 + \frac12r_0k^2 - \frac14P_0  k^4\,,
\label{eq;fs}
\eea
by using the values of the effective range parameters obtained in 
Table \ref{table;s-wave_effective_ranges},
as functions of $T_\alpha$
where $k=\sqrt{1.5\mu T_\alpha}$.
One can see that the fitted curves 
almost overlap at $T_\alpha \simeq 3$-5~MeV,
except for the curve of $S0$,
and when one extrapolates the curves to the lower energies,  
they are scattered.
The curves of $f_s(k)$ 
decreases, is almost the same, and increases at $T_\alpha \simeq 0.4$~MeV,
compared to the values
of the function at $T_\alpha\simeq 3$~MeV, 
depending on the choice of the data sets, $S1$, $S2$, and $S3$, 
respectively.
Thus we find a significant uncertainty in the extrapolation 
to the energy, $T_\alpha\simeq 0.4$~MeV, corresponding to $T_G$. 
We note that the curve of $S3$ vanishes at a small value of $T_\alpha$.
That indicates the presence of the resonant state at the very low energy 
and thus the parameter set $S3$ should be excluded.

\vskip 2mm \noindent
{\bf 4.2. $l=1$ channel}

Three sets of the experimental data for the $p$-wave
phase shift to fit the effective range parameters
are chosen, and labeled as $P0$, $P1$, and $P2$.
$P0$ denotes a data set of the $p$-wave phase shift at $T_\alpha
\simeq 1.5$-3.1~MeV from Table 2 in the Plaga {\it et al.}'s 
paper~\cite{petal-npa87}, and 
$P1$ and $P2$ do those at $T_\alpha= 2.6$-3.0~MeV 
and $2.6$-3.1~MeV, respectively, 
from the Tischhauser {\it et al.}'s paper~\cite{tetal-prc09}.
We note that, as mentioned above,
we chose the largest energies of the data sets
less than the resonance energy, $T_\alpha\simeq 3.23$~MeV.
%
\begin{table}[t]
\begin{center}
\begin{tabular}{c|c c c}
               &
$P0$ &  $P1$ & $P2$ \\ \hline
$\gamma_1$ ($10^3$~MeV$^3$) & 
 $-2.53 \pm 1.09$ & $-3.84\pm 1.27$ & $-4.57\pm0.38$ \\
$r_1$ (fm$^{-1}$) &
 $0.406\pm 0.002$ & $0.405\pm 0.002$ & $0.404\pm 0.000$  \\
$P_1$ (fm) &
 $-0.641\pm0.006$ & $-0.645\pm0.007$ & $-0.649\pm0.002$ 
\\ \hline 
\end{tabular}
\caption{
Fitted values of $p$-wave effective range parameters using three
sets of the experimental data labeled by $P0$, $P1$, and $P2$.
See the text for details.
}
\label{table;p-wave_effective_ranges}
\end{center}
\end{table}

In Table \ref{table;p-wave_effective_ranges} fitted values
of the $p$-wave effective range parameters by using the three sets of
the data labeled by $P0$, $P1$, and $P2$ are displayed.
We find in Table \ref{table;p-wave_effective_ranges} 
the similar tendency to what we found 
in the fitted values of the $s$-wave effective range parameters 
in Table \ref{table;s-wave_effective_ranges}.
The fitted values of $\gamma_1$ are quite sensitive to 
the choice of the data sets, whereas those of $r_1$ and $P_1$ do not.
We can see that the significant cancellations between 
the $r_1$ ($P_1$) term and the term proportional to $k^2$ ($k^4$)
obtained from the $2\kappa k^2(1+\eta^2)h(\eta)$ term 
in Eq.~(\ref{eq;l=1}) where we have $\frac13\kappa \simeq 0.413$~fm$^{-1}$ 
corresponding to the $r_1$ term and 
$-11/15\kappa\simeq - 0.591$~fm corresponding to the $P_1$ term.
We also find that the series expansion of the effective range parameters 
converges at very low energies, up to about $T_\alpha \simeq 1.5$~MeV,
and the significant cancellations occur between the 
terms being proportional to $k^2$ and $k^4$ in the energy range,
$T_\alpha \simeq 2$-3~MeV.

\begin{figure}[t]
\begin{center}
\includegraphics[width=12cm]{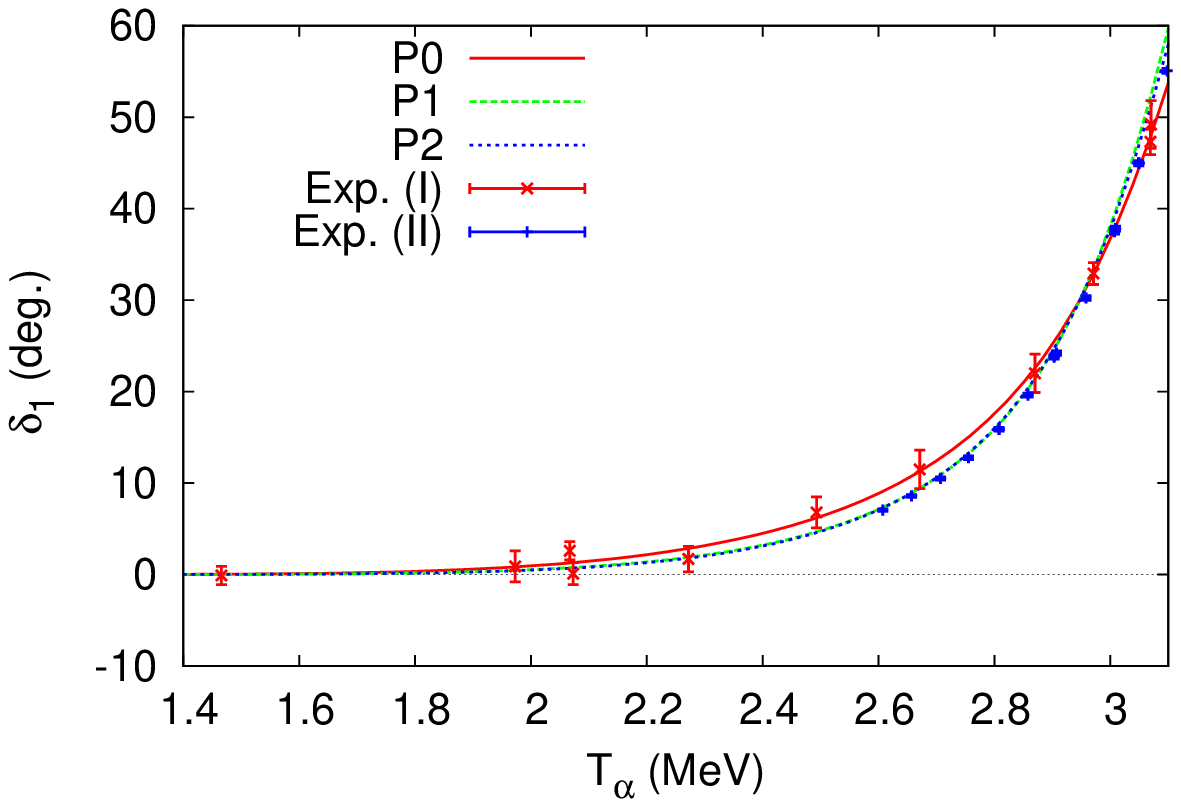}
\caption{
Phase shift of elastic $\alpha$-$^{12}$C scattering for $l=1$, 
$\delta_1$ (deg.), as functions of $T_\alpha$ (MeV).
Three curves are plotted by using three sets of fitted effective range 
parameters (labeled by $P0$, $P1$, $P2$) obtained 
in Table~\ref{table;p-wave_effective_ranges}. Experimental data,
Exp. (I) from Plaga {\it et al.}~\cite{petal-npa87} 
and Exp. (II) from Tischhauser {\it et al.}~\cite{tetal-prc09}, 
are also displayed.
}
\label{fig;phaseshift-pw}
\end{center}
\end{figure}
\begin{figure}[t]
\begin{center}
\includegraphics[width=12cm]{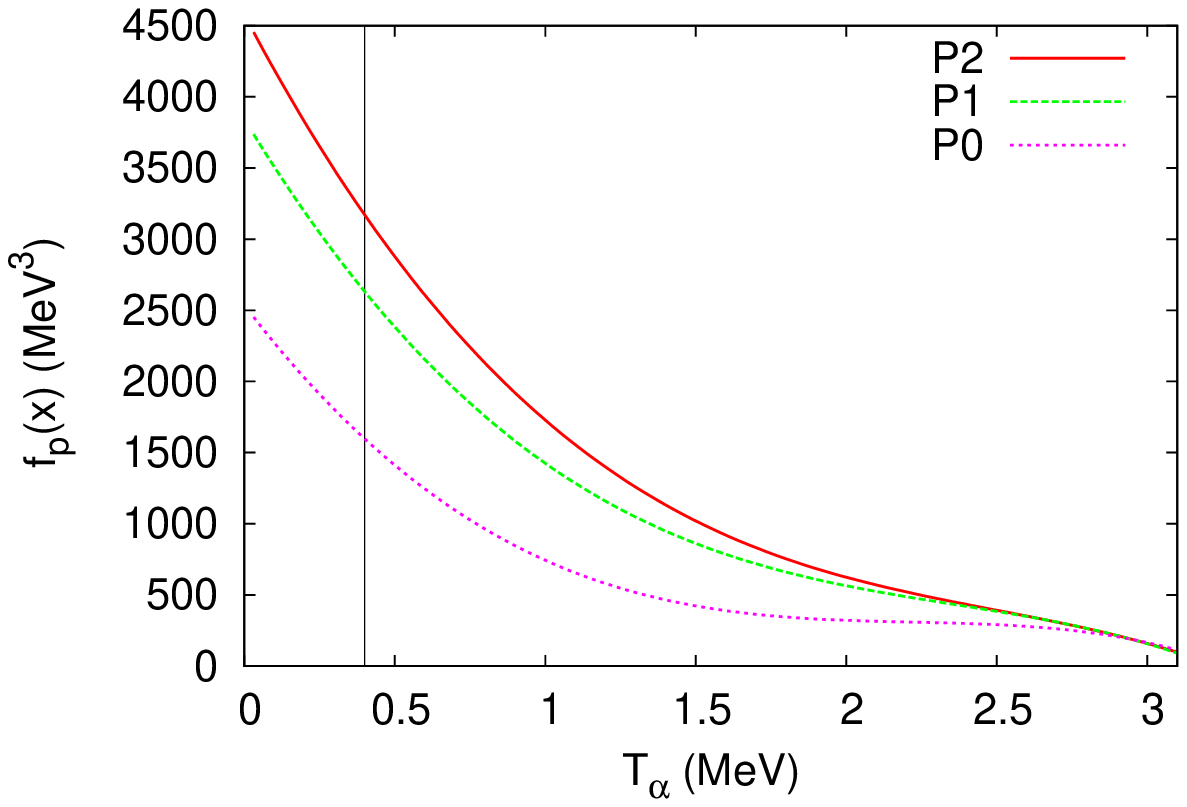}
\caption{
Function $f_p(k)$ defined in Eq.~(\ref{eq;fp}) as functions of $T_\alpha$ (MeV).
Curves are plotted 
by using three sets of values of effective range parameters (labeled 
by $P0$, $P1$, $P2$) obtained 
in Table \ref{table;p-wave_effective_ranges}. 
A vertical line at $T_\alpha=0.4$~MeV is also included.
}
\label{fig;fx-pw}
\end{center}
\end{figure}

In Fig.~\ref{fig;phaseshift-pw}, curves of the phase shift
of the elastic $p$-wave $\alpha$-${}^{12}$C scattering are plotted
by using the fitted effective range parameters 
in Table \ref{table;p-wave_effective_ranges}.
The experimental data are also included in the figure.
We find that the curves well reproduce the data in the energy ranges
where the effective range parameters have been fitted. 

In Fig.~\ref{fig;fx-pw} we plot 
the real part of the denominator of the $p$-wave 
scattering amplitude in Eq.~(\ref{eq;A1}), 
\bea
f_p(k) = - 2\kappa k^2(1+\eta^2) h(\eta) 
- \gamma_1 + \frac12r_1k^2 - \frac14P_1  k^4\,,
\label{eq;fp}
\eea
by using the values of the effective range parameters obtained in 
Table \ref{table;p-wave_effective_ranges} 
as functions of $T_\alpha$. 
One can see that the values of the function $f_p(k)$ are small at
the energy range, $T_\alpha=2.6$-3.0~MeV. In that energy region,
as mentioned above, a significant cancelation among the terms of the 
effective range expansion occurs. 
While the values of the $f_p(k)$ function 
become large
when it is extrapolated to $T_\alpha = 0.4$~MeV, 
due to the relatively large contribution from the $\gamma_1$ term
compared to the other effective range terms of $r_1$ and $P_1$
when the corrections from the $2\kappa k^2(1+\eta^2)h(\eta)$ term
are included. 
This implies that 
because the function $f_p(k)$ appears in the 
denominator of the scattering amplitude
the scattering amplitude is rather suppressed
at $T_G$. 
Thus we cannot qualitatively reproduce the enhancement
of the $S$-factor for the E1 channel,
reported, e.g., in Ref.~\cite{ktf-npa74},
in the extrapolation of the $p$-wave scattering amplitude to $T_G$.

\vskip 2mm \noindent
{\bf 4.3. $l=2$ channel}

Three sets of the experimental data
for the $d$-wave phase shift to fit the effective range parameters
are chosen, and 
labeled 
as $D0$, $D1$, and $D2$.
$D0$ denotes a data set of the $d$-wave phase shift at
$T_\alpha \simeq 1.47$-3.57~MeV from Table 2 
in the Plaga {\it et al.}'s paper~\cite{petal-npa87},
and $D1$ and $D2$ do those at $T_\alpha= 2.6$-3.0 and 2.6-3.4~MeV,
respectively, from the Tischhauser {\it et al.}'s paper~\cite{tetal-prc09}.
We note that 
the maximum energies of the data sets
are chosen as less than the resonance energy, $T_\alpha \simeq 3.57$~MeV. 

\begin{table}[t]
\begin{center}
\begin{tabular}{c|c c c}
               &
$D0$ &  $D1$ & $D2$ \\ \hline
$r_2$ (fm$^{-3}$) &
 $0.37\pm 0.79$ & $0.536\pm 0.001$ & $0.533\pm 0.001$  \\
$P_2$ (fm$^{-1}$) &
 $-6.2\pm 4.3$ & $-5.505\pm0.008$ & $-5.526\pm0.004$ 
\\ \hline 
\end{tabular}
\caption{
Fitted values of $d$-wave effective range parameters using three
sets of the experimental data labeled by $D0$, $D1$, and $D2$.
See the text for details.
}
\label{table;d-wave_effective_ranges}
\end{center}
\end{table}

In Table \ref{table;d-wave_effective_ranges}
fitted values of the $d$-wave effective range parameters 
by using the three data sets introduced above
are displayed. As mentioned before we have chosen $\gamma_2=0$ 
due to the small values of the phase shift 
in those data sets.
We find that large error bars of the fitted parameters from
the data set $D0$ compared to those from the data sets,
$D1$ and $D2$.
We also find the common tendency that the significant cancellations 
between the $r_2$ ($P_2$) term and the term proportional to $k^2$ ($k^4$) 
obtained from the $2\kappa k^4(4+\eta^2)(1+\eta^2)h(\eta)$ term in 
Eq.~(\ref{eq;l=2}) where we have $\frac13\kappa^3\simeq 0.6361$~fm$^{-3}$ 
and $-\frac{51}{15}\kappa \simeq -4.217$~fm$^{-1}$ 
corresponding to $r_2$ 
and $P_2$, respectively. 
For the convergence of the effective range expansion,
we find that there is no convergence for the terms.
There are large cancellations between the terms 
proportional to $k^2$ and $k^4$ at the energies $T_\alpha\simeq 1.5$-2~MeV.
At the larger energies, the $k^4$ term becomes dominant and is significantly
cancelled with the other terms.

\begin{figure}[th]
\begin{center}
\includegraphics[width=12cm]{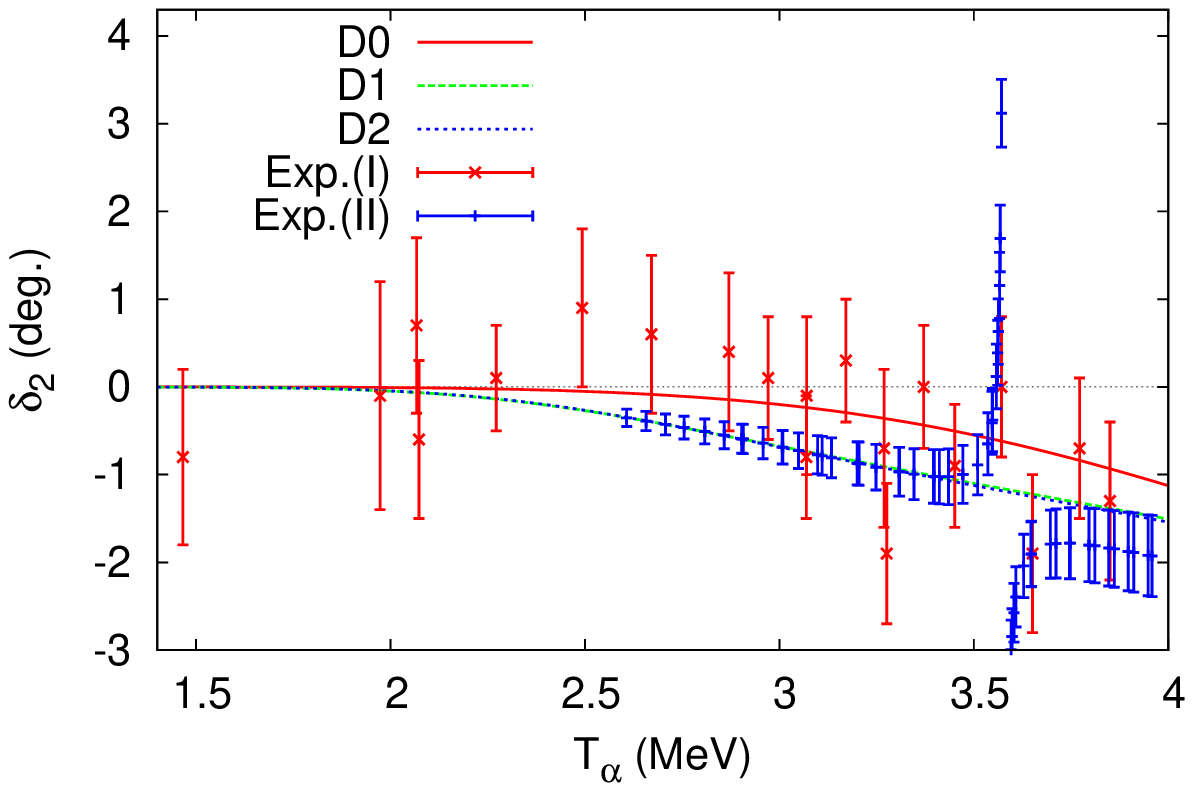}
\caption{
Phase shift of elastic $\alpha$-$^{12}$C scattering for $l=2$, 
$\delta_2$ (deg.), as functions of $T_\alpha$ (MeV).
Three curves are plotted by using three sets of fitted effective range 
parameters (labeled by $D0$, $D1$, $D2$) obtained 
in Table~\ref{table;d-wave_effective_ranges}. Experimental data,
Exp. (I) from Plaga {\it et al.}~\cite{petal-npa87} 
and Exp. (II) from Tischhauser {\it et al.}~\cite{tetal-prc09}, 
are also displayed.
}
\label{fig;phaseshift-dw}
\end{center}
\end{figure}
\begin{figure}[th]
\begin{center}
\includegraphics[width=12cm]{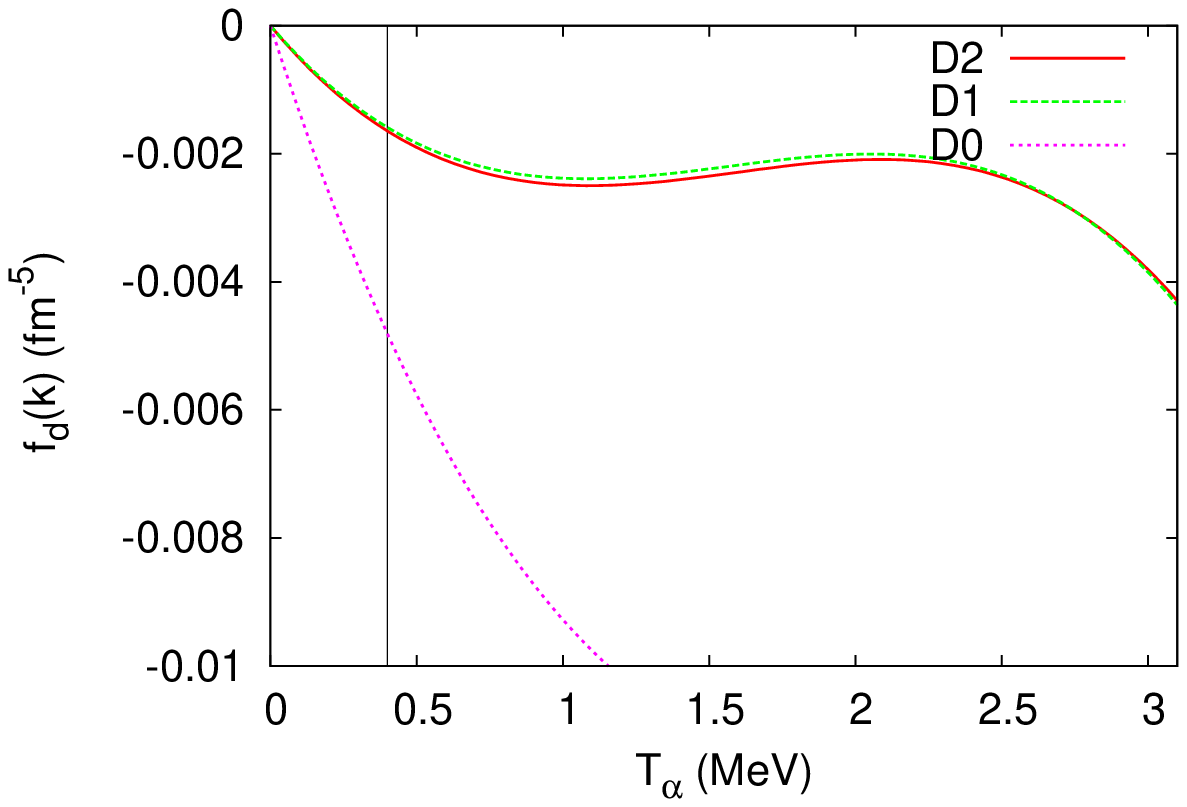}
\caption{
Function $f_d(k)$ defined in Eq.~(\ref{eq;fd}) as functions of $T_\alpha$ (MeV).
Curves are plotted 
by using three sets of values of effective range parameters (labeled 
by $D0$, $D1$, $D2$) obtained 
in Table \ref{table;d-wave_effective_ranges}. 
A vertical line at $T_\alpha=0.4$~MeV is also included.
}
\label{fig;fx-dw}
\end{center}
\end{figure}

In Fig.~\ref{fig;phaseshift-dw}, curves of the $d$-wave phase shift
are plotted
by using the fitted values of the effective range parameters obtained
in Table \ref{table;d-wave_effective_ranges}. 
The experimental data are also included in the figure.
We can see that the curves plotted in the figure well reproduce the 
experimental data in the energy range below the resonance energy, 
$T_\alpha\simeq3.57$~MeV, 
and the error bars of the Tischhauser {\it et al.}'s data 
are significantly smaller than those of the Plaga {\it et al.}'s data.  

In Fig.~\ref{fig;fx-dw} we plot curves of 
the real part of the denominator of the $d$-wave scattering
amplitude in Eq.~(\ref{eq;A2}), 
\bea
f_d(k) = - 2\kappa k^4(4+\eta^2)(1+\eta^2) h(\eta) 
- \gamma_2 + \frac12r_2k^2 - \frac14P_2  k^4\,,
\label{eq;fd}
\eea
by using the values of the effective range parameters 
in Table \ref{table;d-wave_effective_ranges} 
as functions of $T_\alpha$.
One can see that the curves vanish in the limit where $k\to 0$
because of the assumption of $\gamma_2=0$. 
In addition, 
the tracks of the extrapolation for the data sets of $D1$ and $D2$
are quite different from that of $D0$, whereas
as mentioned before, the curve from the $D0$ data set should have 
a large uncertainty due to the large error bars of the fitted 
parameters. 
Furthermore, we find that the curves 
of $D1$ and $D2$ are almost
flat from $T_\alpha\simeq 2$~MeV to $T_\alpha\simeq 0.4$~MeV, 
so the extrapolated cross section would be almost the same 
except for the common factor $C_\eta^2$ at $T_G$ 
and $T\simeq 2$~MeV. 
Thus we cannot qualitatively reproduce the enhancement
of the $S$-factor for the E2 channel either, reported, e.g., in
Ref.~\cite{lk-npa83}, in the extrapolation 
of the $d$-wave scattering amplitude to $T_G$.

\vskip 2mm \noindent
{\bf 5. Discussion and conclusions}

In this work we introduce an EFT for 
the $^{12}$C($\alpha$,$\gamma$)$^{16}$O at $T_G$ 
where the $\alpha$ and $^{12}$C states are treated as the 
elementary-like cluster fields,
and apply the theory to the study of the phase sifts of the 
elastic $\alpha$-$^{12}$C scattering for $l=0,1,2$ channels
at the low energies.
The expression of the scattering amplitudes for $l=0,1,2$ channels 
is obtained
in terms of the three effective range parameters. 
The effective range parameters are fitted by using the sets of
the experimental data in the energy ranges below the 
resonance energies in which 
the phase shifts are smoothly varying.
In the parameter fitting we find that it is 
difficult to incorporate 
the pole structure for 
the subthreshold $1_1^-$ and $2_1^+$ states 
in the amplitudes.
Nevertheless we find that the experimental data at the low energies
are well reproduced by the curves plotted using 
the fitted effective range parameters.
To qualitatively study the uncertainty of 
the extrapolation to $T_G$
due to the fitting of the parameters to the experimental phase shift data,
we extrapolate the real part of the denominator of the
scattering amplitudes to $T_G$,
and find that there
are the significant uncertainties.
Because parameters deduced from the scattering
phase shifts for $l=1$ and 2 channels 
may play important roles 
in the extrapolation of the $S$-factors of the radiative capture reaction 
for the E1 and E2 transitions,
we discuss our results in the parameter fitting and 
the extrapolation of the scattering amplitudes 
to $T_G$ in some details below.

In the parameter fitting for the $l=1$ channel, 
the phase shift data entirely appear as the low energy tail of 
the $1_2^-$ resonant state, and 
the tail from the $1_1^-$ bound state
at $T=-0.045$~MeV can be scarcely seen in the data.
As we have found above, 
to reproduce the data
it is not necessary to include the 
subthreshold bound state.
In addition, the amplitude extrapolated to $T_G$ is largely suppressed. 
Those observations in fact have repeatedly been pointed out 
in the literatures~\cite{h-npa97,h-npa01,dds-prc10}. 
Indeed, to estimate the radiative capture $E1$ transition cross section
at $T_G$, 
it would be essential to include an explicit degree of 
freedom for the $1_1^-$ state in the theory,
possibly along with the $1_2^-$ state~\cite{g-prc09}.
Although the tail of the the $1_1^-$ state is not clearly seen in
the elastic scattering data,
the significance of the $1_1^-$ bound state 
may be seen in the other experimental data 
such as the $\beta$-delayed $\alpha$ decay from $^{16}$N, 
$^{16}$N($\beta^-$)$^{16}$O($\alpha$)$^{12}$C
\cite{hhw-prl70,betal-prl93,tetal-prc10},
whose minimum energy is $T = 0.6$~MeV, 
and the radiative $E1$ capture cross section whose minimum energy
now becomes less than $T=1$~MeV, 
see, e.g., Refs.~\cite{retal-epja99,aetal-prc06,petal-prc12}.  

In the parameter fitting for the $l=2$ channel,
the phase shift data at $T_\alpha=1.5$-3.5~MeV 
show a down slope up to the $2_2^+$ resonant state appearing at 
$T_\alpha=3.57$~MeV.
As demonstrated above, it is not difficult to fit 
the restricted data by using the three effective range parameters,
but it appears not easy to precisely decompose it 
into three ingredients, the tail of the subthreshold state, 
that of the resonant state, and a background.  
Moreover, as discussed, it is not easy to include 
the $2_1^+$ subthreshold state at $T=-0.24$~MeV 
because of the feature of the scattering amplitudes
(represented in terms of the effective range parameters)
in the present study.  
In the extrapolation, as seen above, we find that 
almost the similar magnitude of the denominator of the scatter amplitude 
at $T_G$ and $T_\alpha\simeq 2$~MeV.
This observation can be questionable because 
the series of the effective range expansion 
we obtained does not converge.
Thus to extrapolate the radiative $E2$ capture cross section to $T_G$
in the present theory,
it would be essential to include an explicit degree of freedom for
the subthreshold $2_1^+$ state as well.
Moreover, as pointed out 
by Sparenberg~\cite{s-prc04}, 
asymptotic normalization constant (ANC)
for the $2_1^+$ state is not possible to determine from single 
channel scattering phase shift data, 
it may be necessary to fix couplings for the $2_1^+$ state 
from model calculations, such as 
a supersymmetric potential model assuming rotational band
for the $0_2^+$, $2_1^+$, $4_1^+$, and $6_1^+$ states 
for $^{16}$O~\cite{s-prc04}, 
or a microscopic cluster model~\cite{dd-prc08}. 
It may also be possible to estimate the parameters
from other experimental data, such as a cascade transition to the 
$2_1^+$ state~\cite{b-prc01,bk-ajp91} 
or $\gamma$ distribution of the radiative capture
rate at the very low energies~\cite{retal-epja99,aetal-prc06,petal-prc12}.

\vskip 2mm \noindent
{\bf Acknowledgements}

The author would like to thank 
K. Kubodera for encouragement to the present work and 
S.-W. Hong and T.-S. Park for discussions.
This work is supported by the Sunmoon University Research Grant of 2015.

\end{document}